\documentclass[conference]{IEEEtran}
\IEEEoverridecommandlockouts
\usepackage{cite}
\usepackage{amsmath,amssymb,amsfonts}
\usepackage{algorithmic}
\usepackage{graphicx}
\usepackage{textcomp}
\usepackage{xcolor}
\usepackage{hyperref}
\def\BibTeX{{\rm B\kern-.05em{\sc i\kern-.025em b}\kern-.08em
    T\kern-.1667em\lower.7ex\hbox{E}\kern-.125emX}}
\begin{document}

\title{FEAD: Figma-Enhanced App Design Framework for Improving UI/UX in Educational App Development}

\author{\IEEEauthorblockN{Tianyi Huang}
\IEEEauthorblockA{\textit{Student Ambassador} \\
\textit{App Inventor Foundation}\\
Fremont, California, USA \\
tianyi@appinventorfoundation.org}
}
\maketitle

\begin{abstract}

Designing user-centric mobile applications is increasingly essential in educational technology. However, platforms like MIT App Inventor—one of the world's largest educational app development tools—face inherent limitations in supporting modern UI/UX design. This study introduces the Figma-Enhanced App Design (FEAD) Method, a structured framework that integrates Figma's advanced design tools into MIT App Inventor using an identify-design-implement workflow. Leveraging principles such as the 8-point grid system and Gestalt laws of perception, the FEAD Method empowers users to address design gaps, creating visually appealing, functional, and accessible applications. A comparative evaluation revealed that 61.2\% of participants perceived FEAD-enhanced designs as on par with professional apps, compared to just 8.2\% for baseline designs. These findings highlight the potential of bridging design with development platforms to enhance app creation, offering a scalable framework for students to master both functional and aesthetic design principles and excel in shaping the future of user-centric technology.

\end{abstract}


\section{Introduction}

The proliferation of mobile applications has fundamentally transformed global communication, information access, and daily life. With over five billion smartphone users worldwide and approximately 1,300 new apps released daily, the demand for innovative and user-centric applications is at an all-time high \cite{StatApp, StatPhone}. In this digital age, tools like MIT App Inventor play an important role in empowering the next generation of creators \cite{wolber2015}. As a free, open-source platform, MIT App Inventor allows users—regardless of technical background—to design fully functional mobile applications using an intuitive drag-and-drop interface \cite{MITAppInventor}. With over 11 million users across 200 countries and regions, the platform has bridged the gap between computational thinking and real-world innovation \cite{Lao2023}.

However, despite its accessibility and educational impact, MIT App Inventor faces limitations in its UI/UX design capabilities. Users often encounter a rigid grid system, limited design customization, and alignment inconsistencies across devices, which restrict the creation of modern and visually appealing apps. These shortcomings diminish the overall user experience and cause obstacles for users trying to meet professional design standards \cite{UICommunity}. Feedback from student communities and educational organizations—such as App-In Club, a global nonprofit that supports student-driven app development—highlights the urgent need for enhanced design capabilities within the platform \cite{AppInClub}.

To address these challenges, leveraging external design tools into the MIT App Inventor workflow offers a promising solution. Figma, a web-based design platform popular for its collaborative features and comprehensive design functionalities, provides tools such as vector editing, flexible grids, and a vast library of design components \cite{Figma}. By incorporating Figma’s capabilities, developers can potentially overcome MIT App Inventor’s inherent design constraints, crafting apps that are both visually modern and functional.

This study introduces the Figma-Enhanced App Design (FEAD) Method, a complete approach to integrating Figma into the MIT App Inventor development process to enhance UI/UX design. By combining Figma’s design tools with fundamental design principles—such as the 8-point grid system and Gestalt laws of perception—this method directly addresses the platform’s limitations while elevating the app development experience. The findings demonstrate that the FEAD Method also serves as a scalable framework for incorporating professional design tools into educational programming environments. This study provides a detailed guide for educators, students, and developers, enabling them to apply the FEAD Method to unlock greater creativity, usability, and impact in app development.

\section{Related Works}

\subsection{Design Extensions in MIT App Inventor}

Efforts to improve MIT App Inventor’s design capabilities have included extensions like UI Enhancer and MakeViewUp, which introduce additional customization options such as gradients, padding, and ripple effects \cite{UIEnhancer, MakeViewUp}. While these extensions provide helpful tools for refining app aesthetics, they are constrained by the limitations of MIT App Inventor’s component library. Moreover, they come with a major limitation: apps built with extensions can’t be published to the MIT App Inventor gallery, restricting their accessibility. This constraint limits the usability of such enhancements for users who wish to showcase their apps broadly. Addressing this limitation requires an alternative approach that bypasses the reliance on extensions while offering the same design flexibility.

\subsection{External Design Tools in Education}

The use of external design tools such as Adobe XD, Sketch, and Figma have been lauded for their ability to streamline collaborative workflows and produce professional-grade designs \cite{Coursera}. Studies exploring their integration into educational settings have demonstrated significant improvements in both engagement and output quality \cite{Wang2022}. For instance, in higher education, researchers have found that incorporating Figma into app development courses helps students grasp the importance of design principles while promoting collaboration \cite{Borysova2024}. However, these studies often focus on advanced techniques or the selection of optimal external design tools, leaving their integration with programming platforms like MIT App Inventor still underexplored. To address this gap, the FEAD Method was developed with the aim of mitigating the divide between design tools and programming platforms.

\subsection{Impact of Design on Learning}

Research in human-computer interaction (HCI) and educational technology highlights the importance of intuitive UI/UX design. Poorly designed interfaces can increase cognitive load and frustration, particularly for beginners, while well-designed apps promote ease of use and continued engagement \cite{Hollender2010, Banu2021, Peters2018}. This study builds on existing findings that demonstrate the dual role of design as both a functional tool and a motivational driver in education, reinforcing the purpose and development of the FEAD Method.

\begin{figure}[!htb]
    \centering
    \includegraphics[width=\linewidth]{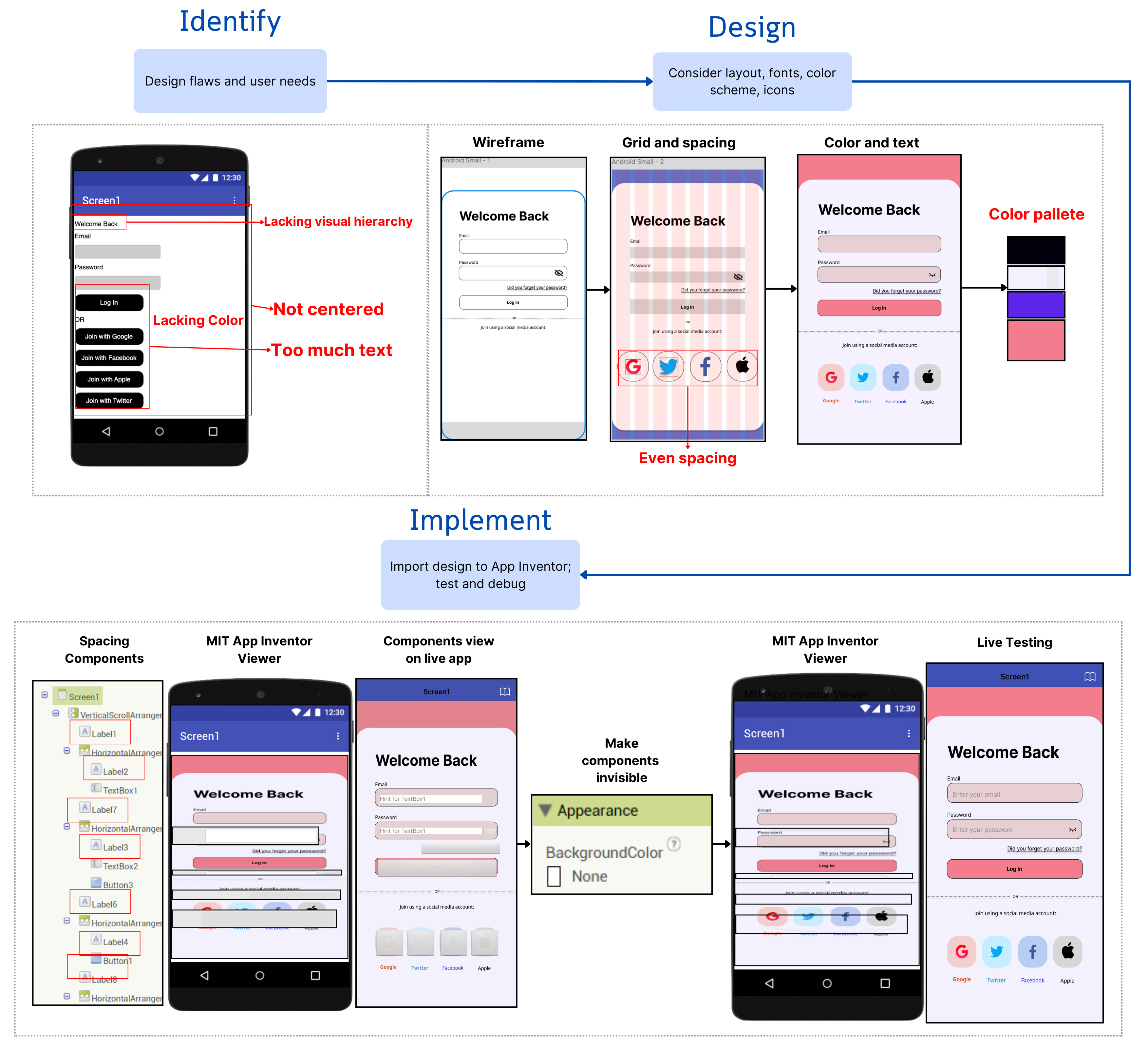}
    \caption{This diagram illustrates the FEAD methodology applied to a login screen. (1) Identify: Initial design flaws, such as poor visual hierarchy, misaligned elements, and excessive text, are pinpointed. (2) Design: Solutions are shown using wireframes, refined with grids for even spacing, and polished with a color palette to improve aesthetics. (3) Implement: The finalized design is deployed in MIT App Inventor, where components are arranged and tested to create a fully functional and visually polished app.}
    \label{fig:fead_methodology}
\end{figure}

\section{Methodology}

\subsection{Identify}

We selected an existing shopping list application from the MIT App Inventor gallery as the example for our improvements \cite{Gallery}. This application allowed users to add items via typing, remove individual entries, and clear the entire list. An initial usability analysis was conducted to identify UI/UX shortcomings. The analysis revealed several issues: a cluttered layout, weak visual hierarchy, and inadequate color contrast. These issues were assessed based on Gestalt principles of perception \cite{Wagemans2012, Wertheimer1928} and established UI design guidelines \cite{Johnson2010}.

The cluttered layout resulted from the close proximity of functionally distinct components (e.g., add, remove, and clear buttons), which could lead to user confusion and increased cognitive load. The weak visual hierarchy made it difficult for users to intuitively understand the primary functions of the app, potentially hindering efficient task completion. Inadequate color contrast posed accessibility concerns, particularly for users with visual impairments, contravening the Web Content Accessibility Guidelines (WCAG) \cite{W3C}.

\begin{figure}[!htb]
    \centering
    \includegraphics[width=0.3\linewidth]{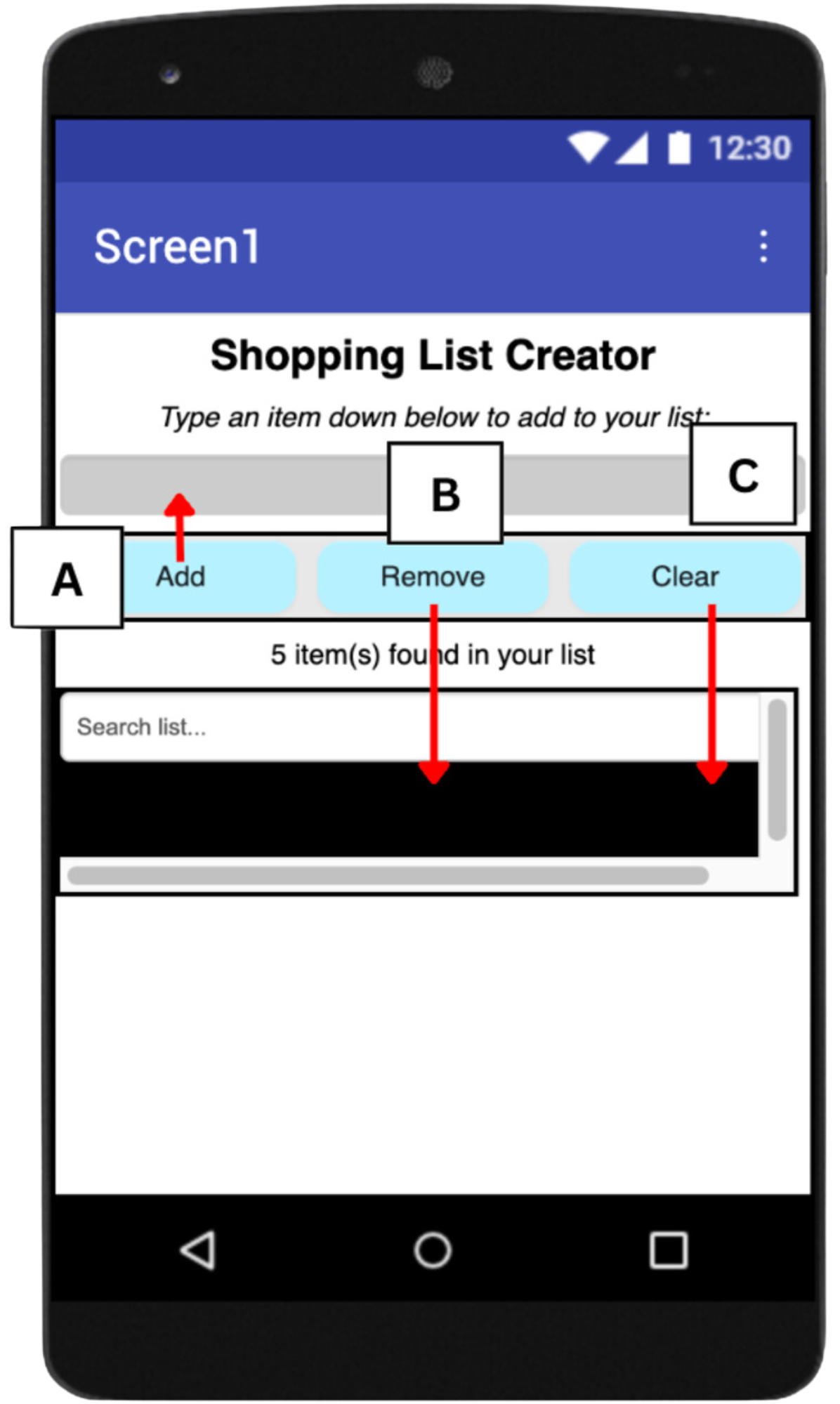}
    \caption{The original app design is displayed, highlighting components A (Add), B (Remove), and C (Clear). The close proximity of these components and the lack of visual distinction emphasize the need for a redesign to enhance usability and aesthetics.}
    \label{fig:original_design}
\end{figure}

\subsection{Design}

\subsubsection{Wireframing}

To address the identified issues, we developed wireframes to conceptualize the improved UI layout. Wireframing is an important step in the design process, allowing designers to focus on content placement, functionality, and user flow without the distraction of visual design elements. We utilized Figma, a professional design tool, to create the wireframes and applied an 8-point grid system to ensure consistent spacing and alignment.

The wireframes were designed following Gestalt principles, specifically the laws of proximity and common region, which state that elements close to each other or within the same bounded area are perceived as related \cite{Palmer1992}. By reorganizing input elements (e.g., the input field and related labels) and output elements (e.g., list view and action buttons) into distinct regions, we aimed to enhance the logical flow and reduce cognitive load.

Standard icons were used in place of text labels to improve clarity and universal understanding, adhering to best practices in iconography \cite{Lidwell2010}. The use of icons also contributed to a cleaner visual aesthetic and streamlined the user experience.

\begin{figure}[!htb]
    \centering
    \includegraphics[width=0.8\linewidth]{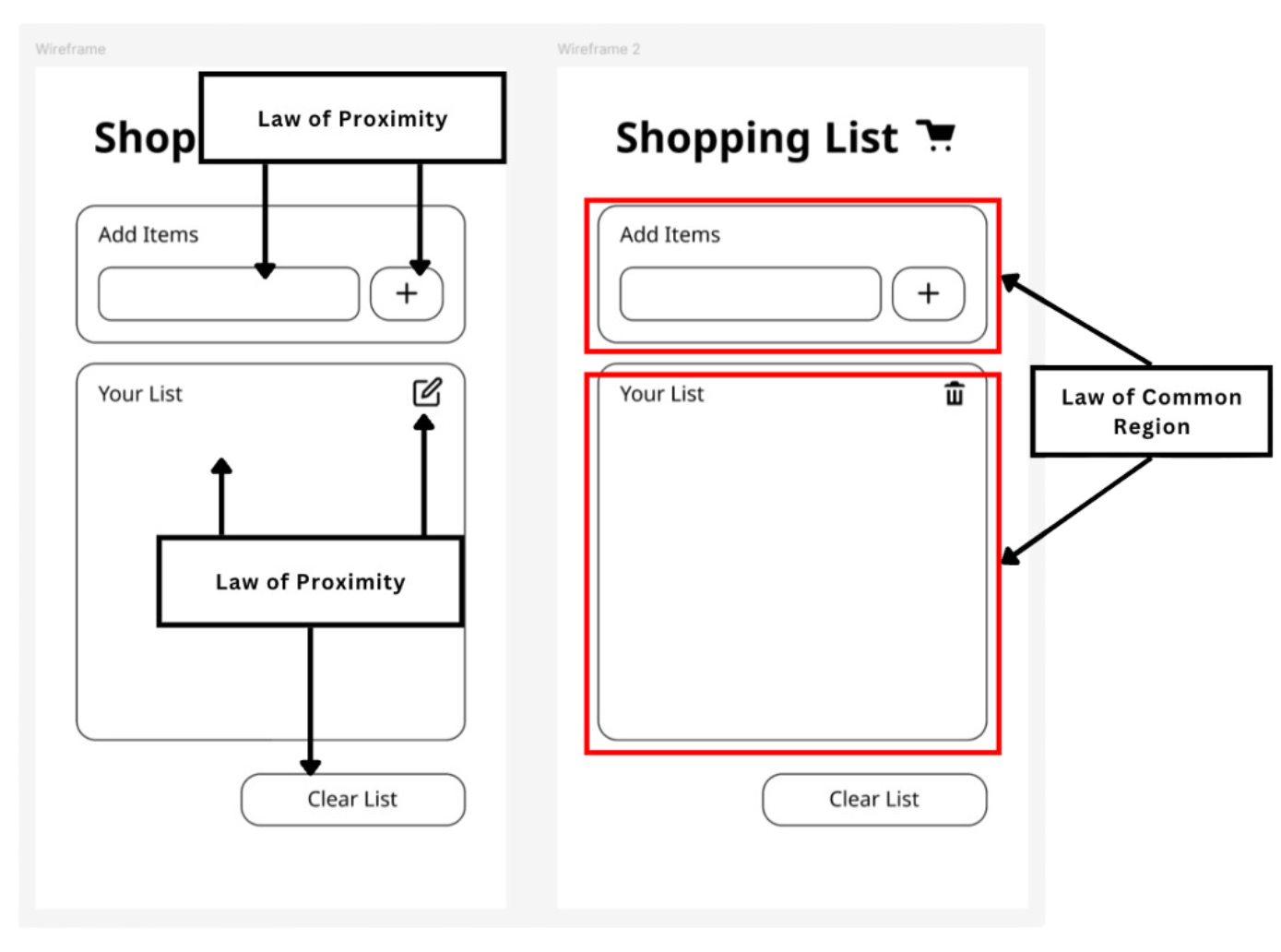}
    \caption{This improved wireframe demonstrates the logical repositioning of components to enhance usability. The left screen applies the Law of Proximity by grouping related elements, while the right screen incorporates the Law of Common Region to further organize the interface with distinct visual boundaries.}
    \label{fig:improved_wireframe}
\end{figure}

\begin{figure}[!htb]
    \centering
    \includegraphics[width=0.8\linewidth]{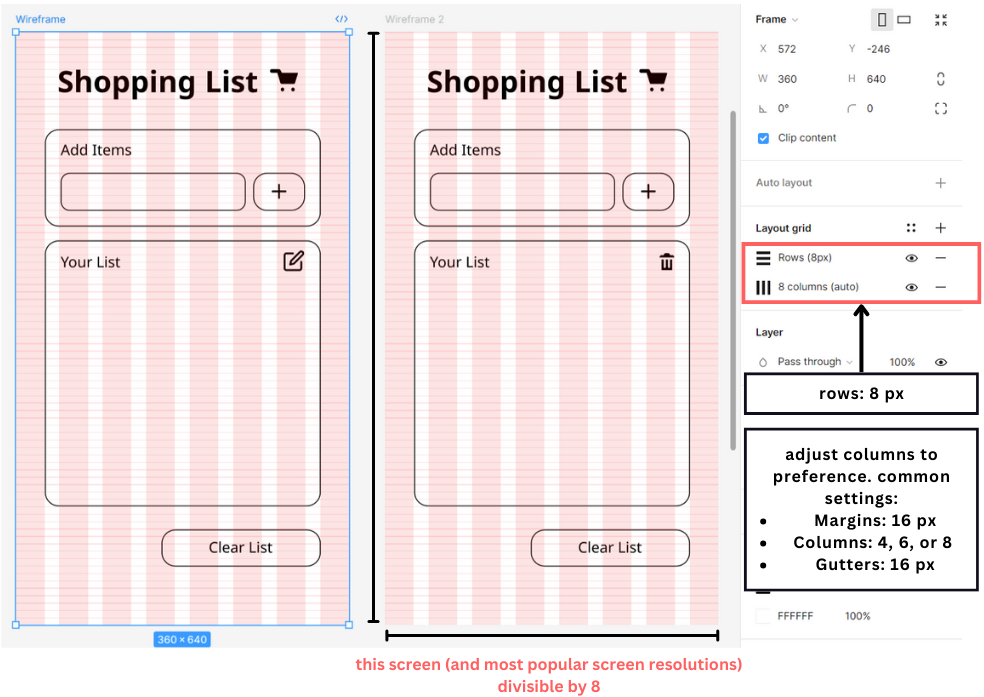}
    \caption{The spacing of the design follows the 8-point grid system through the layout grid feature in Figma. Rows are set to a height of 8 px, with standard settings for margins and gutters set at 16 px, allowing for consistent alignment across the interface.}
    \label{fig:point_grid}
\end{figure}

\begin{figure}[!htb]
    \centering
    \includegraphics[width=0.7\linewidth]{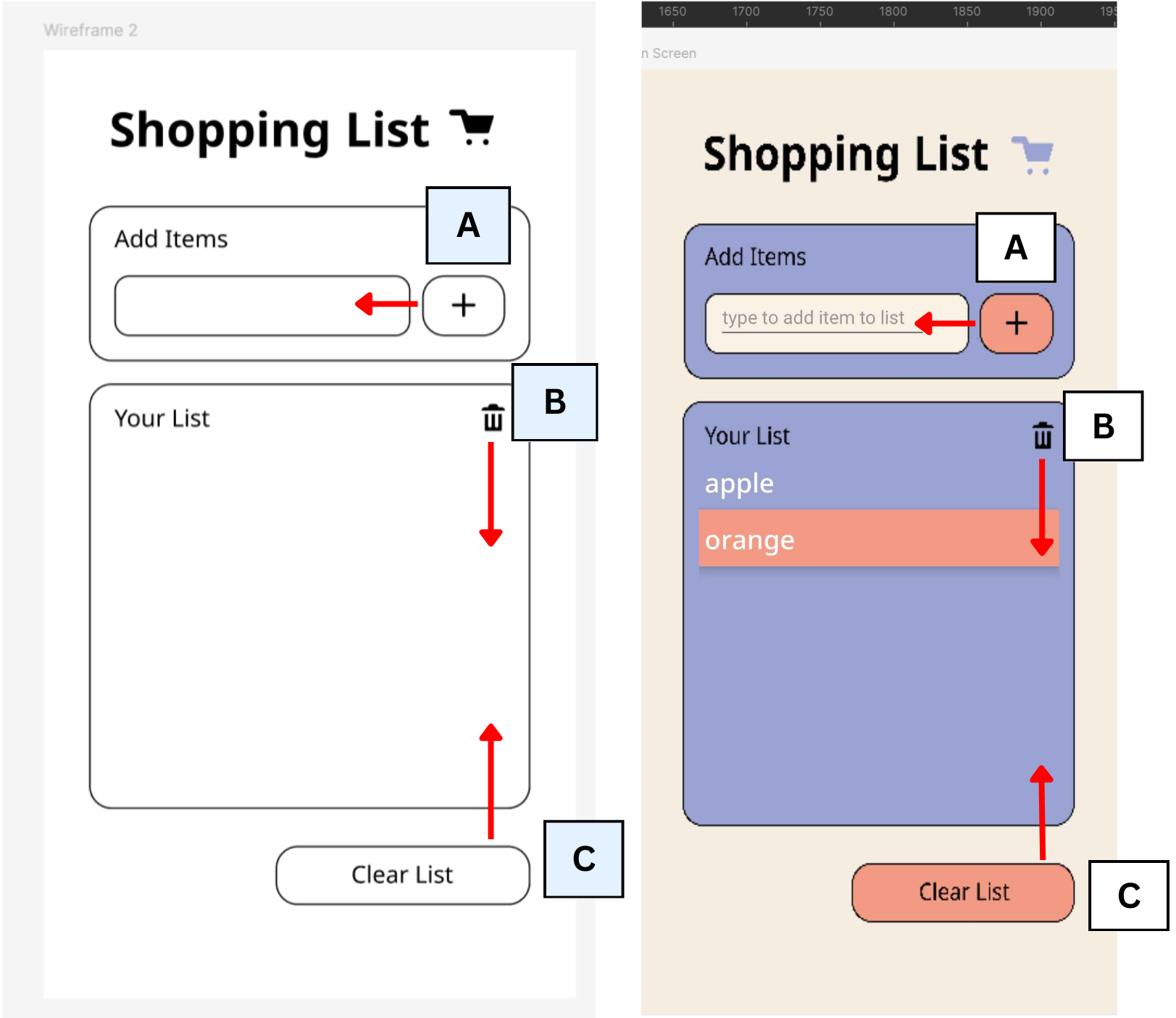}
    \caption{In the improved wireframe design, components A (Add), B (Remove), and C (Clear) are redesigned for clarity. They are separated, replaced with icons, and positioned closer to their respective elements for better visual grouping and usability, as pointed out by the red arrows.}
    \label{fig:redesigned_wireframe}
\end{figure}

\subsubsection{Design Development}

Building upon the wireframes, we developed the final screen designs with a focus on visual aesthetics. We adopted the 60-30-10 rule for color scheme composition, a guideline used in interior design and UI/UX design to create balanced and harmonious color palettes \cite{Gordon2021}. In our design, beige was selected as the dominant color (60\%), blue as the secondary color (30\%), and orange as the accent color (10\%) to draw attention to important interactive elements.

To ensure color choices met accessibility standards, we leveraged Realtime Colors and other contrast-checking tools to evaluate the color combinations in real-time \cite{realtime}. These tools assessed the contrast ratios between background and text, adhering to WCAG 2.1 guidelines, which recommend a minimum contrast ratio of 4.5:1 for normal text and 3:1 for large text \cite{W3C}. Our chosen color palette achieved contrast ratios exceeding 7:1, satisfying Level AAA compliance for optimal readability.

\begin{figure}[!htb]
    \centering
    \includegraphics[width=0.9\linewidth]{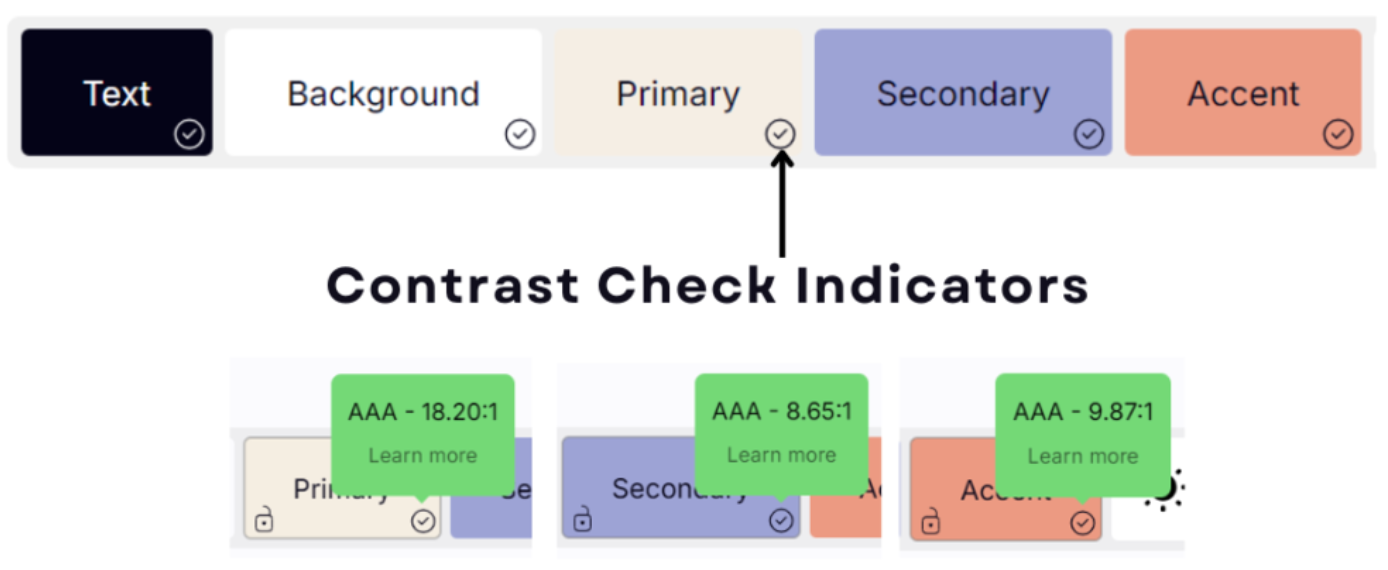}
    \caption{This displays the results of the color contrast check, demonstrating that all text and background color combinations meet accessibility standards, with AAA compliance indicated for each combination.}
    \label{fig:color_contrast}
\end{figure}

The final design also considered typography, iconography, and visual hierarchy. We selected fonts that were legible and ensured that interactive elements were visually distinct and intuitive to use \cite{Bringhurst2016}. 

\subsection{Implementation}

\begin{figure}[!htb]
    \centering
    \includegraphics[width=\linewidth]{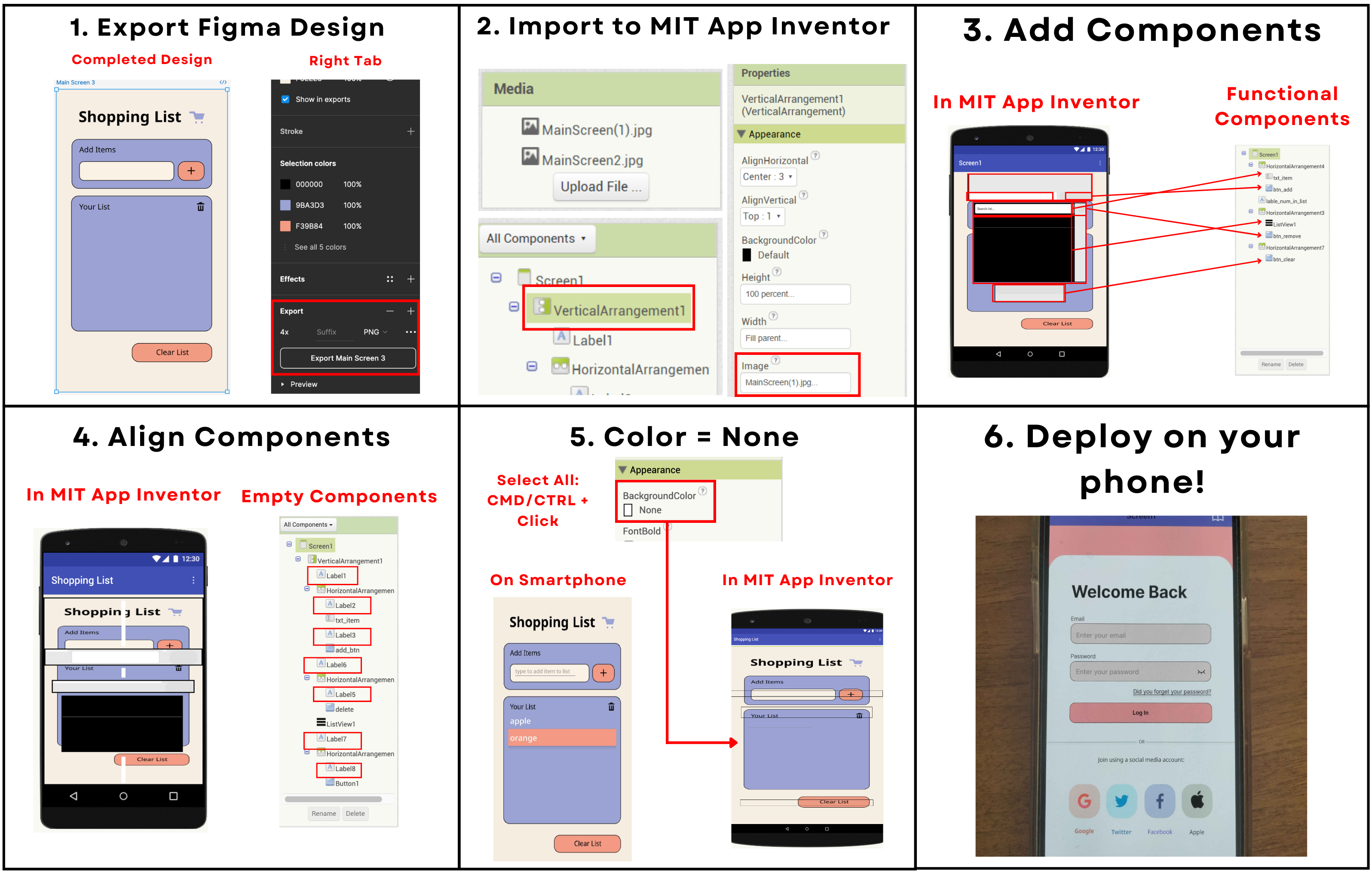}
    \caption{Implementation process for integrating Figma designs into MIT App Inventor: (1) Designs are exported from Figma. (2) The design is imported into MIT App Inventor using a vertical arrangement. (3) Functional components are added to replicate the design. (4) Components are aligned carefully within MIT App Inventor. (5) Components are set to invisible by setting the background color to 'None.' (6) The final app is deployed and runs on a smartphone.}
    \label{fig:implementation_process}
\end{figure}

We imported the designs from Figma into App Inventor by exporting the visual assets and integrating them within the application's interface using vertical arrangements. To maintain the visual integrity of the design, we set the background color of components to "none," making them invisible and allowing the imported design elements to serve as the primary visual components. Interactive elements such as buttons and text fields were overlaid onto the visual design using empty labels and arrangements to preserve functionality without disrupting the aesthetic.

Live testing was conducted using the MIT App Inventor Companion app, which allowed real-time previewing and adjustment of the components on an actual device. This step was necessary for identifying and resolving alignment issues due to differences in device dimensions and screen sizes. 

To enhance interactivity and provide immediate feedback to users, we implemented image swaps and visual cues within the app's code. For instance, buttons were programmed to change appearance when pressed, reinforcing the action taken by the user. Additionally, we matched the accent colors of interactive components, such as list picker selections, to the established color scheme to maintain visual consistency.

After ensuring all components functioned as intended and the UI aligned correctly across devices, we finalized the implementation by setting the background colors of all components to "none," allowing the designed visuals to remain unobstructed.

\section{Results}

\subsection{Quantitative Analysis}
To evaluate the impact of integrating Figma into the UI/UX design process within MIT App Inventor, we conducted an anonymous survey involving 50 participants, particularly high school students who have created apps with MIT App Inventor. Participants were presented with two app designs: one created solely using MIT App Inventor (baseline) and another designed with Figma in MIT App Inventor (FEAD).

Participants rated each design on several aspects, including overall UI/UX experience and color scheme, using a scale from -1 to 1 (where -1 indicates a negative perception, 0 is neutral, and 1 is positive).

The quantitative analysis revealed a clear preference for the Figma-enhanced design, indicating a significant improvement in user perception when Figma is incorporated into the design process.

\begin{figure}[htbp]
\centerline{\includegraphics[width=\columnwidth]{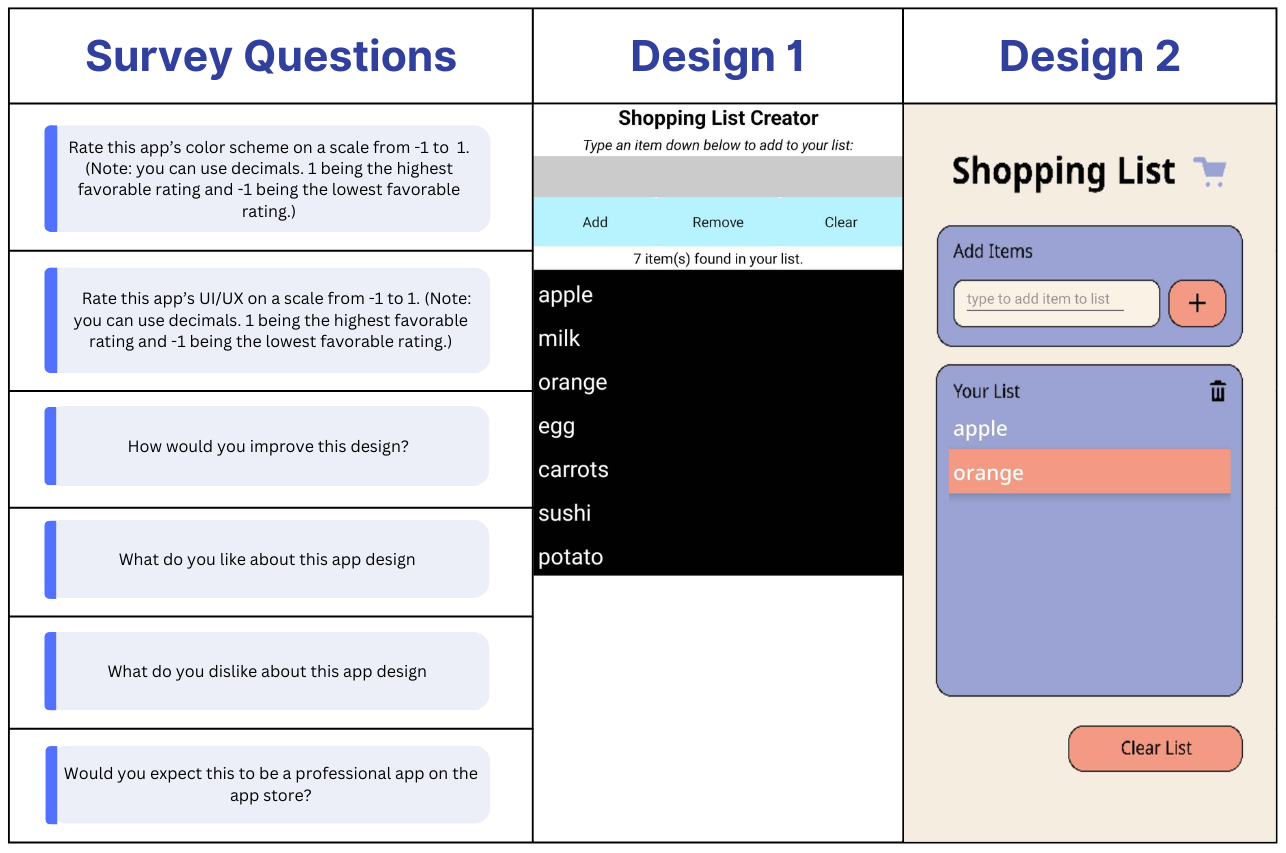}}
\caption{Survey questions and designs compared. Each question was presented twice to participants, once for Design 1 (baseline) and once for Design 2 (redesigned with FEAD), allowing for direct comparison of user feedback on color scheme, UI/UX, strengths, weaknesses, and perceived professionalism.}
\label{fig_survey}
\end{figure}

\begin{figure}[htbp]
\centerline{\includegraphics[width=0.9\columnwidth]{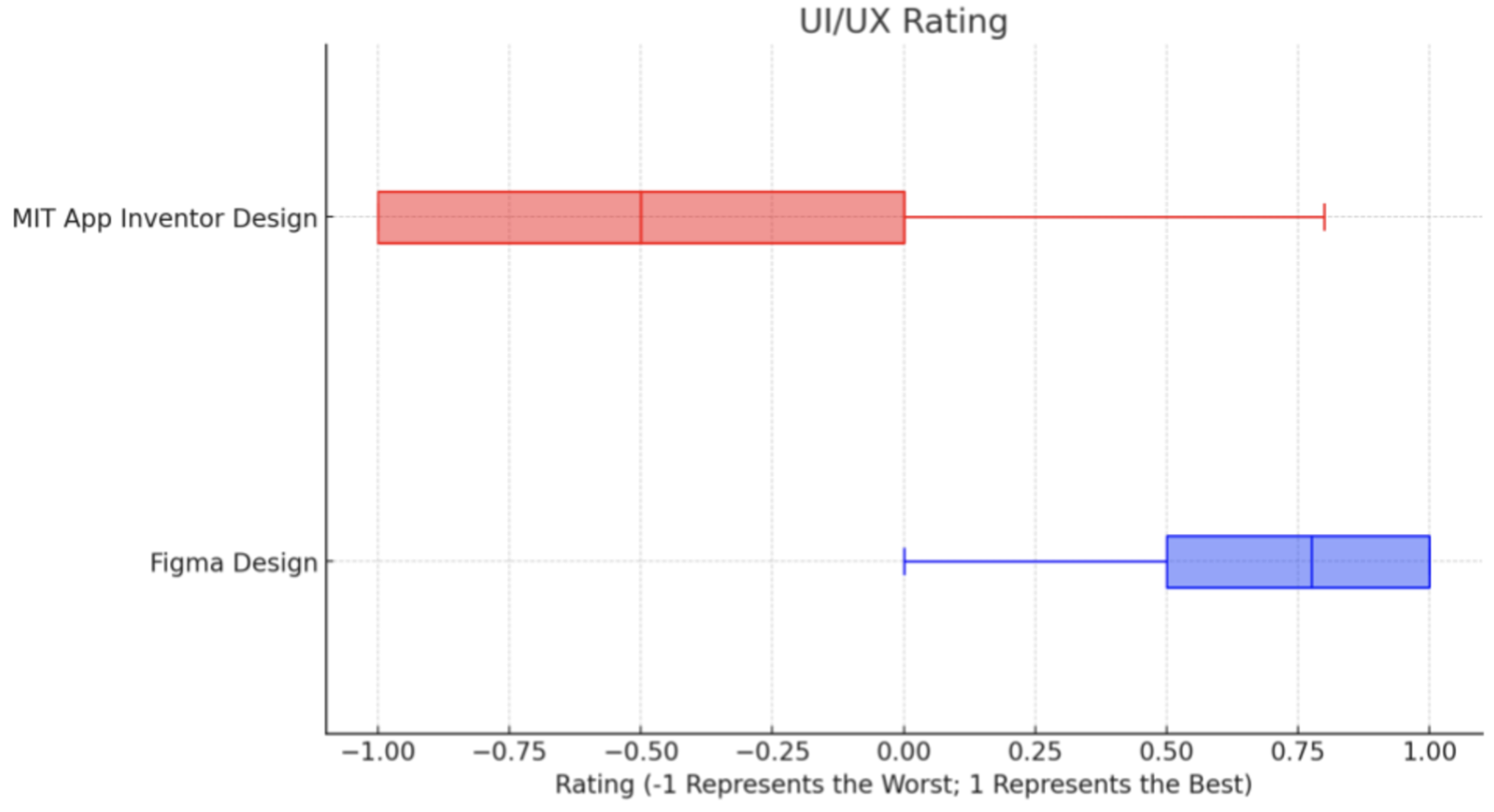}}
\caption{Comparison of UI/UX ratings between the MIT App Inventor design (baseline) and the Figma design (FEAD), showing a significant improvement in user experience for the Figma-enhanced design.}
\label{fig_uiux_ratings}
\end{figure}

\begin{figure}[htbp]
\centerline{\includegraphics[width=0.9\columnwidth]{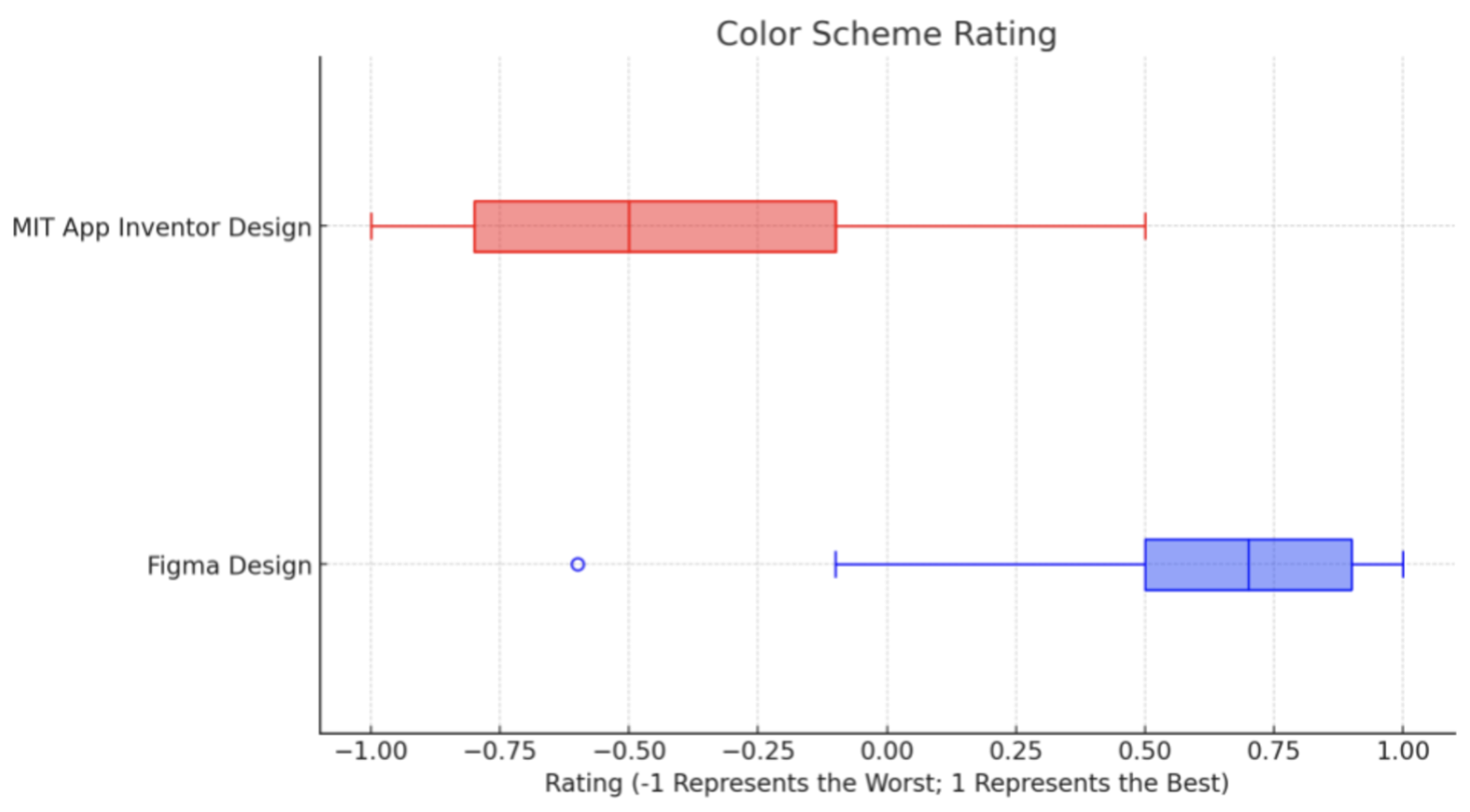}}
\caption{Comparison of color scheme ratings between the MIT App Inventor design (baseline) and the Figma design (FEAD), showing a higher aesthetic appeal of the Figma-enhanced design.}
\label{fig_color_ratings}
\end{figure}

\begin{itemize}
    \item Mean score for overall UI/UX experience:
    \begin{itemize}
        \item Figma design: 0.727
        \item MIT App Inventor design: -0.380
    \end{itemize}
    \textit{Interpretation:} Figma-enhanced design scored significantly higher, proving its superior user experience.
    \item Mean score for color scheme:
    \begin{itemize}
        \item Figma design: 0.719
        \item MIT App Inventor design: -0.423
    \end{itemize}
    \textit{Interpretation:} Figma-enhanced design outperformed MIT App Inventor, proving its visual appeal.
\end{itemize}

\subsection{Qualitative Analysis}
In addition to quantitative ratings, participants provided open-ended feedback on their experiences with each design. We performed a thematic analysis of the responses, extracting key descriptors and sentiments associated with each design. The findings are visualized in the following word clouds.

\begin{figure}[htbp]
\centerline{\includegraphics[width=0.9\columnwidth]{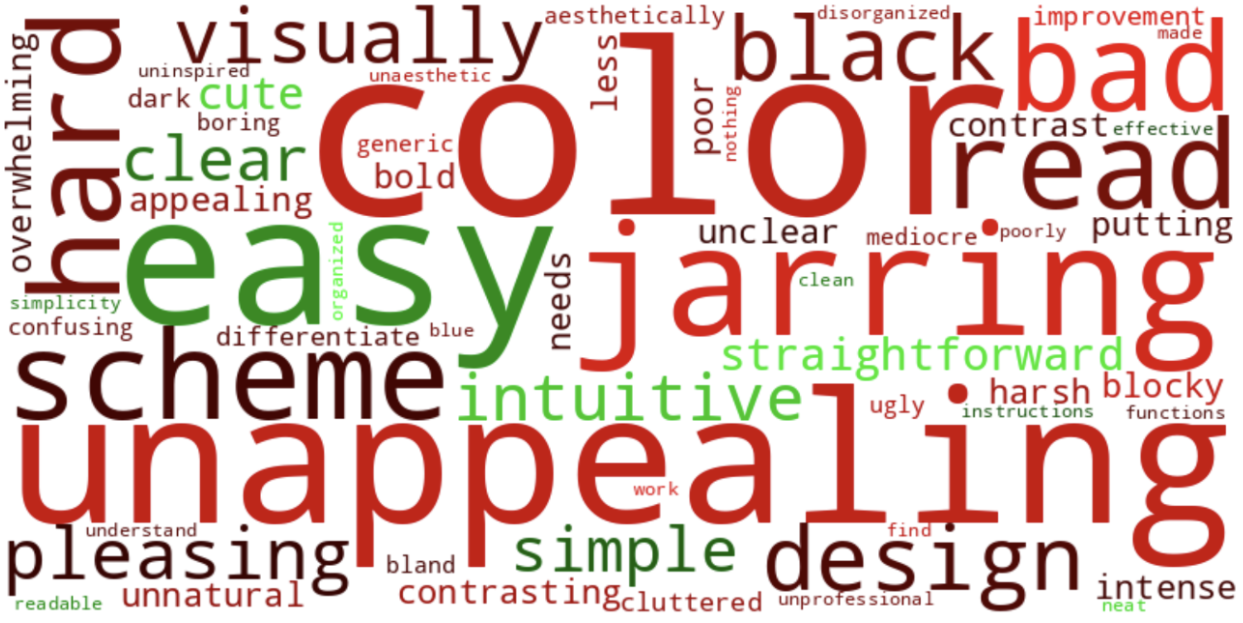}}
\caption{Word cloud illustrating user feedback on the original MIT App Inventor design (baseline).}
\label{fig_wordcloud_baseline}
\end{figure}

The word cloud for the purely MIT App Inventor design is dominated by negative terms such as "unnatural," "jarring," "unappealing," and "hard," reflecting users' dissatisfaction with both the aesthetic and functional aspects of the design. The prevalence of these terms reveals the limitations of MIT App Inventor's native design capabilities.

\begin{figure}[htbp]
\centerline{\includegraphics[width=\columnwidth]{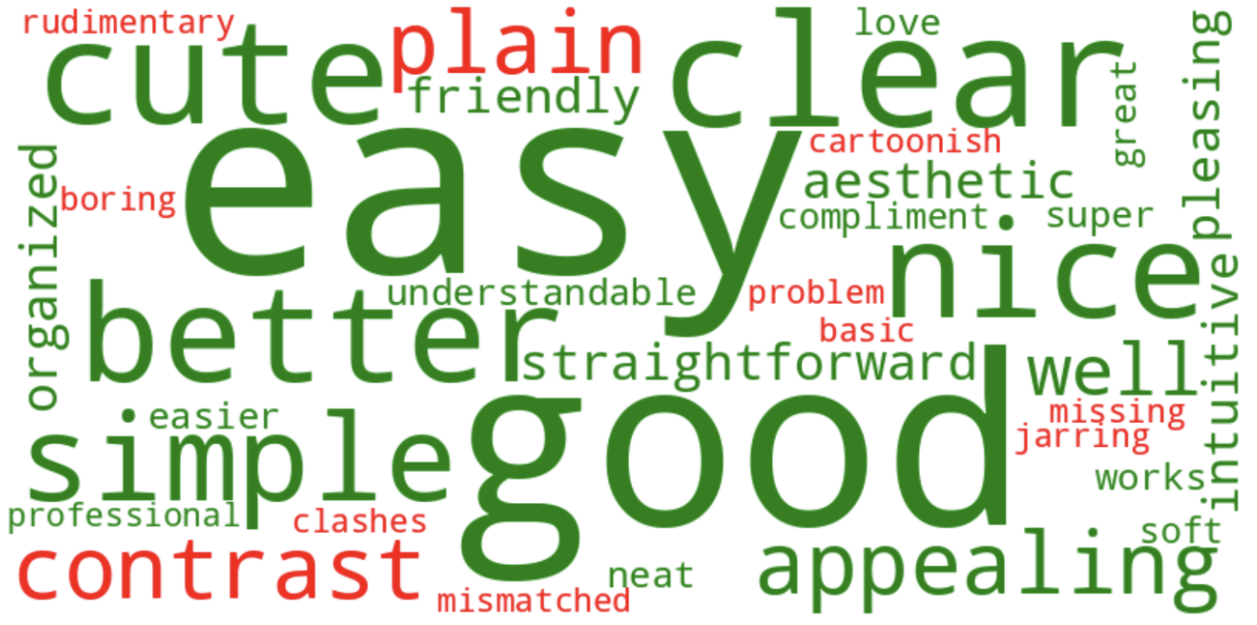}}
\caption{Word cloud illustrating user feedback on the Figma-enhanced design (FEAD).}
\label{fig_wordcloud_fead}
\end{figure}

Conversely, the word cloud for the Figma-enhanced design features positive descriptors like "aesthetic," "intuitive," "clear," and "pleasing." These terms indicate that users found the Figma-enhanced design more visually appealing and user-friendly, aligning with modern UI/UX standards.

\subsection{Perceived Professionalism}
Participants were also asked to identify which design they perceived as originating from a professional app. A significant majority, 61.2\%, selected the Figma-enhanced design, while only 8.2\% chose the MIT App Inventor design. The remaining participants were either undecided or felt that neither design appeared professional. This disparity demonstrates the impact of professional design tools on user perception of app quality.

\subsection{Implications}
Overall, the results demonstrate that incorporating external design tools like Figma can significantly improve the UI/UX quality of apps developed with MIT App Inventor. The enhanced designs meet more users' aesthetic expectations and also improve usability and overall satisfaction. These improvements are critical in educational contexts, where the goal is to create engaging and effective learning experiences.

The substantial differences in user ratings suggest that the limitations of MIT App Inventor's native design capabilities can be effectively addressed by integrating modern design principles through external tools. This approach allows developers and educators to leverage the simplicity and accessibility of MIT App Inventor while delivering a user experience that aligns with modern UI/UX standards.

\section{Limitations and Future Work}

\subsection{Limitations}
\begin{itemize}
    \item \textbf{Screen Size Adaptability:} Integrating Figma designs into MIT App Inventor poses challenges in achieving consistent user interfaces across different screen sizes. Without precise adjustments for each device, UI elements may become misaligned or improperly scaled, adversely affecting the user experience and necessitating testing and modification.
    \item \textbf{Restricted Components:} MIT App Inventor's limited customization for interactive components can cause restrictions that prevent users from fully implementing specific layouts and advanced UI concepts envisioned in Figma.
    \item \textbf{Limited Interactivity:} Currently, Figma designs can only be imported as static background images within MIT App Inventor. To add functionality, developers must overlay invisible components and manually align them with the design elements. This process is time-consuming and complicates the development workflow.
\end{itemize}

\subsection{Future Work}
\begin{itemize}
    \item \textbf{Developing a Custom AI Alignment Tool:} A major future direction can be creating a custom AI-powered alignment tool specific to MIT App Inventor \cite{openai2024}. This tool would analyze the layout differences between the design viewed on the desktop (in MIT App Inventor) and on a mobile device, providing precise guidance on the adjustments needed to achieve perfect alignment. By automating this process, developers would save significant time and effort.
    \item \textbf{Figma Templates for MIT App Inventor:} To lower the barrier to entry for educators and users, another key direction is to develop a library of high-quality, customizable Figma templates specifically designed for integration with MIT App Inventor. These templates would follow best practices in UI/UX design and address common needs in educational and general app development. Offering ready-made templates would enable users to focus on functionality while ensuring a polished and professional aesthetic.
    \item \textbf{Pilot Testing and Surveys:} Building on the initial user surveys, a more extensive pilot testing program is necessary to validate the integration method. Users would create complete example apps using both methods—direct design in MIT App Inventor and Figma-enhanced workflows—and provide detailed feedback on their experiences. By analyzing their ratings and feedback, we can gain a better understanding of the method’s strengths, weaknesses, and overall user satisfaction. This data would also inform future iterations of the integration workflow and tools.
\end{itemize}

\section{Ethical Considerations}
This study involved an anonymous survey to evaluate the impact of integrating external design tools, such as Figma, into the MIT App Inventor workflow. All participants were high school students familiar with app development and voluntarily participated in the survey after being informed of the study’s purpose and methodology.

The survey was conducted anonymously to ensure participant privacy. No personal or identifiable information was collected during the process, and responses were used solely for the purposes of this study. Participants were made aware of their rights to withdraw from the survey at any time without penalty. These measures align with ethical standards for research involving human participants and prioritize transparency, respect, and confidentiality.

\section{Conclusion}
This study successfully introduced the Figma-Enhanced App Design (FEAD) Method to address the UI/UX limitations of MIT App Inventor by integrating external design principles and tools. Through a combination of Figma and established design frameworks, we demonstrated a measurable improvement in user experience and aesthetic quality. Apps developed using the FEAD Method achieved mean scores of 0.727 for overall user experience and 0.719 for color scheme ratings, significantly outperforming the baseline designs, which only scored -0.380 and -0.423.

By merging accessibility with professional-grade design, the FEAD Method not only enhances the visual appeal of MIT App Inventor apps but also sets a precedent for integrating external tools into educational programming environments. This framework paves the way for broader adoption of design-first approaches in app development education, promoting creativity, engagement, and innovation among students and educators alike.

\section{Acknowledgments}
I would like to express my gratitude to my mentor, Dr. Natalie Lao, for her encouragement and support in this project.

I would also like to acknowledge the members of the App-In Club Research Team, as well as my mentees—Zehan Li, Siyuan Yang, and Sarina Feng—for their time and assistance in creating the figures, distributing the surveys, and reviewing the paper.

\bibliographystyle{IEEEtran}
\bibliography{ms}

\end{document}